\documentclass[%
 reprint,
 amsmath,amssymb,
 aps,
]{revtex4-2}

\usepackage[normalem]{ulem}
\usepackage{graphicx}% Include figure files
\usepackage{dcolumn}% Align table columns on decimal point
\usepackage{bm}% bold math
\usepackage{cancel}
\usepackage{nicefrac}
\usepackage{multirow}
\usepackage[dvipsnames]{xcolor} % for colors 
\usepackage{xspace}
\usepackage[hidelinks]{hyperref}
\usepackage{caption}
\usepackage{subcaption}
\usepackage{comment}

\newcommand{\arctanh}[0]{\text{arctanh}}

\begin{document}

\preprint{APS/123-QED}

\title{Thermodynamic Geometry Through Second Order Phase Transitions}

\author{Omer M. Basri}
 \email{omer.basri@weizmann.ac.il}
\author{Oren Raz}
 \email{oren.raz@weizmann.ac.il}
\affiliation{%
 Department of Physics of Complex Systems,\\Weizmann Institute of Science, Rehovot 76100, Israel. 
}%

\date{\today}

\begin{abstract}
A common approach to quantify excess dissipation in slowly driven thermodynamic processes is through the use of a Riemannian metric on the space of control parameters, where optimal driving protocols follow geodesics.
Near phase transitions, this geometric picture breaks down as the metric diverges and geodesics may cease to exist.
Using Widom scaling, we analyze this framework for several universality classes and show that in some cases the thermodynamic length across the phase transition remains finite.
We then demonstrate a numerical approach for computing minimal paths in such systems. We show that, in some regimes, the shortest path crosses the phase transition -- even when alternative paths confined to a single phase exist.
\end{abstract}

\maketitle

\paragraph{Introduction}
Since the earliest developments of equilibrium statistical mechanics to modern nonequilibrium statistical mechanics there have been attempts to cast the theory in a geometric form \cite{gibbs1873method,ruppeinerThermodynamicsRiemannianGeometric1979, weinholdMetricGeometryEquilibrium1975,crooksMeasuringThermodynamicLength2007,janyszekGeometricalStructureState1989,PhysRevE.111.034113,raz2016geometric,van2024geometric,brandner2020thermodynamic,RevModPhys.67.605, mandalAnalysisSlowTransitions2016,zhong2024beyond}. 
One of the most useful approaches is to regard the excess dissipation as a Riemannian metric. This idea was introduced already in the 80's \cite{salamonThermodynamicLengthDissipated1983,salamonRelationEntropyEnergy1984}, following purely geometrical concepts suggested already in the 70's \cite{ruppeinerThermodynamicsRiemannianGeometric1979, weinholdMetricGeometryEquilibrium1975}. It was based on two assumptions: (i) endoreversibility---the system remains in internal equilibrium throughout the process, though not necessarily in equilibrium with the environment; (ii) the dissipation is only a function of the system's mean response time. The latter assumption restricts the formalism’s applicability to systems with nearly constant relaxation time. Despite these limitations, the framework has proven valuable for identifying minimal-entropy-production protocols in processes such as chemical reactions and distillation \cite{Andresen2000,andresenCurrentTrendsFiniteTime2011}. 

The two limiting assumptions were lifted by using a closely related metric, which only assumes linear response theory to be applicable \cite{sivakThermodynamicMetricsOptimal2012, zulkowski2012geometry,rotskoffDynamicRiemannianGeometry2015,sivak2016thermodynamic,blaber2020skewed, rotskoffGeometricApproachOptimal2017}. In this formulation, the system’s relaxation dynamics are incorporated into the metric, hence removing the dependence on a constant response-time and eliminating the endoreversible assumption. This generalized framework extends thermodynamic geometry to a much broader class of systems---including those with widely varying time scales and microscopic systems such as biomolecular or single-molecule systems---and has found applications in optimizing computational and physical processes \cite{rotskoffGeometricApproachOptimal2017}. A similar framework was suggested for quantum systems \cite{scandi2019thermodynamic}.

A particularly challenging class of systems exhibiting widely separated time scales are those undergoing phase transitions. The application of thermodynamic geometry to such systems has drawn significant attention \cite{janyszekRiemannianGeometryThermodynamics1989, PhysRevE.51.1006}, including efforts to identify optimal driving protocols across critical regions (e.g., the 2D ferromagnetic Ising model \cite{rotskoffDynamicRiemannianGeometry2015}). In many cases of interest, e.g., antiferromagnets \cite{vivesUnifiedMeanfieldStudy1997}, crossing the phase transition cannot be avoided as the two phases are separated by a phase transition surface. Yet, the divergence of thermodynamic length near phase transitions has received limited study. Notably, a divergence in the metric tensor does not necessarily imply a divergence in thermodynamic length. However, even if the length remains finite, other geometric quantities such as the curvature may diverge, signaling the critical behavior. 

In this paper, we analyze the behavior of the thermodynamic metric near second-order phase transitions. Using the Widom scaling and the dynamical scaling hypothesis, we show that in some class of models, the divergence of the metric at the phase transition does not imply divergence of dissipation when crossing the phase transition. Therefore, the thermodynamic distance between states in the different phases is finite. The class of models with this property includes several mean field models, as well as the Ising model in dimensions 3 and above. To demonstrate our findings, we apply the formalism to the mean field antiferromagnetic Ising model, where  we determine the optimal protocols connecting two points. Surprisingly, we find that in some cases the optimal protocol between two points that are in the same phase nevertheless crosses to the other phase and back.

\paragraph{Framework} \label{sec:framework}
Here we briefly define the key metrics and scaling tools used in our study. The dissipation metric, first introduced in \cite{sivakThermodynamicMetricsOptimal2012}, is defined as
\begin{equation}\label{Eq:Metric}
    g_{\mu\nu} = \beta \int_0^\infty \langle \delta X_\mu(t)\, \delta X_\nu(0) \rangle\, dt =
    \beta \mathcal{T}_{\mu\nu} \frac{\partial^2 \ln Z}{\partial \lambda^\mu \partial \lambda^\nu}
\end{equation}
where $\mathcal{T}_{\mu\nu}$ is the time integrated temporal response matrix, $Z$ the partition function, and $X_\mu$ the conjugate variable to $\lambda^\mu$ -- the natural variables of the free energy, e.g., $\{X_1,X_2\} = \{S, M\}$ for $F(T, h)$ \footnote{We assume that the Hamiltonian of the system is linear with $\lambda_i$.}.
In the linear response regime, the dissipated availability which expresses the excess dissipation in performing the trajectory $\boldsymbol{\lambda(t)}$ in parameter space, is given by
\begin{equation} \label{Eq:DissipationRelation}
    \mathcal{A} = \int_{\boldsymbol{\lambda}(t)} g_{\mu\nu}\, \dot{\lambda}^\mu \dot{\lambda}^\nu\, dt.
\end{equation}
It is related to the length associated with the thermodynamic metric, defined by
\begin{equation}
    \mathcal{L} = \int_{\boldsymbol{\lambda}(s)} \sqrt{g_{\mu\nu}\, \dot{\lambda}^\mu \dot{\lambda}^\nu}\, ds
\end{equation}
through the inequality
\begin{equation}\label{Eq:A_bound}
    \mathcal{A} \ge \frac{\mathcal{L}^2}{T}
\end{equation}
 (here $T$ is the protocol duration, and $\lambda(s)$ parametrize the trajectory in parameter space). This bound is saturated for protocols with $g_{\mu\nu}\, \dot{\lambda}^\mu \dot{\lambda}^\nu=const.$ at all time \cite{salamonThermodynamicLengthDissipated1983}.

%\subsection{Scaling Hypothesis} \label{subsec:ScalingHypothesisFramework}
Near second order phase transitions, second derivatives of the free energy, as well as response times ---and hence the metric defined in Eq.(\ref{Eq:Metric})--- diverge. The asymptotic form of these divergences is determined by critical exponents, that are common to all models that belong to the same universality class. These critical exponents can be analyzed using Widom's scaling \cite{stanley1987introduction}, which for Ising type models relates the free energy $\varphi$ near the critical point $(T_c, H_c)$ to reduced variables, often chosen to be $t = (T - T_c)/T_c$ and $h = (H - H_c)/H_c$. The Widom scaling then states that
\begin{equation} \label{eq:ScalingHyp}
    \varphi \propto t^b f\!\left(\frac{h}{t^\Delta}\right)
\end{equation}
where $f(x)$ is continuous and $f(x) \sim x^p$ as $x \to \infty$ \cite{Kardar_2007}.
The exponents $b$, $\Delta$, $p$, are enough to uniquely dictate the standard critical exponents through the hyperscaling relations, e.g., the heat capacity exponent $\alpha$ defined through $C_H(t)\sim |t|^{-\alpha}$ is given by $\alpha = 2 - b$.
This approach captures divergences of the free energy second derivatives across phase transitions, but does not address relaxation times, which are critical for the dissipation metric in Eq.(\ref{Eq:Metric}). This can nevertheless be done using the dynamical scaling relations \cite{hohenberg1977theory}, that relates the relaxation time $\tau$ and correlation length $\xi$ as
\begin{equation}
    \tau \propto \xi^z
\end{equation}
and the free energy scales as $\varphi \propto \xi^{-d}$ \cite{Kardar_2007, tauberCriticalDynamicsField2014}.
Combining these yields
\begin{equation}\label{eq:TimeResponsScaling}
    \tau \propto \varphi^{-z/d} \propto t^{-bz/d}\, f^{-z/d}\!\left(\frac{h}{t^\Delta}\right)
\end{equation}
linking dynamical and thermodynamic critical behavior. Note that since we use the correlation length in this analysis, it cannot be used for mean field models.

%\subsection{Divergence of Lengths}
With the Widom and dynamical scaling, it is thus possible to find the divergence of the metric near the phase transition of various equivalence classes. Diverging metric, however, does not necessarily imply a divergent thermodynamic length.
Since it is always possible to saturate the equality in  Eq.(\ref{Eq:A_bound}),  finite thermodynamic length guarantees the existence of a finite-dissipation protocol. In addition,  $\mathcal{L} \to \infty$ implies $\mathcal{A} \to \infty$ for all finite time protocols.
Consequently, the convergence of thermodynamic length determines whether finite-dissipation protocols exist under the formalism of thermodynamic geometry. 
If $\dot{\mathcal{L}}(s) = \sqrt{g_{\mu\nu}\dot{\lambda}^\mu\!\left(s\right)\dot{\lambda}^\nu\!\left(s\right)} \propto s^{-a}$ near criticality (as $s \rightarrow 0$), the total length diverges only for $a > 1$. Note, however, that finite distance across a phase transition does not necessarily inform  us about the metric itself, or the intrinsic geometry of the manifold. The metric might diverge for flat space \cite{PhysRev.119.1743,Szekeres:1960gm}, and the manifold might have divergent curvature despite having finite lengths and a non-divergent metric.

\paragraph{Results}

%\subsection{Scaling Hypothesis} \label{subsec:ScalingHypothesisResults}

%\subsubsection{Scaling Hypothesis for the Free Energy}
The metric near a phase transition diverges when either one of the free energy derivatives or the time responses diverge. Since the singular part of the thermodynamic potential scales as Eq.(\ref{eq:ScalingHyp}), the scaling of the derivatives depends on the direction of the approach to criticality. Specifically, the two possible asymptotic behaviors are 
\begin{align}
    \left(\frac{\partial^2 \ln Z}{\partial \lambda^\mu \partial \lambda^\nu}\right) \Big|_{\frac{h}{t^\Delta} \to \text{const.}} &\propto 
    \begin{pmatrix}
        t^{b-2} & t^{b-\Delta-1} \\
        t^{b-\Delta-1} & t^{b-2\Delta}
    \end{pmatrix}, \\[4pt]
    \left(\frac{\partial^2 \ln Z}{\partial \lambda^\mu \partial \lambda^\nu}\right) \Big|_{\frac{h}{t^\Delta} \to \infty} &\propto
    \begin{pmatrix}
        t^{-2}h^p & t^{-1}h^{p-1} \\
        t^{-1}h^{p-1} & h^{p-2}
    \end{pmatrix}.
\end{align}
The time response divergence in the two cases can be obtained from  Eq.(\ref{eq:TimeResponsScaling}). From the static scaling relation, we obtain that the divergence of $\tau$ follows
\begin{equation} \label{Eq:tauPropto}
    \tau \propto
    \begin{cases}
        t^{-bz/d}, & \frac{h}{t^\Delta} \to \text{const.} \\[4pt]
        h^{-pz/d}, & \frac{h}{t^\Delta} \to \infty
    \end{cases}
\end{equation}
Together, the derivatives of the free energy and the response time imply the following scaling: For $\frac{h}{t^\Delta} \to \text{const.}$, one finds that
\begin{align}
    g_{tt}\,\dot{t}^2 &\propto s^{b(1-z/d)-2} \\
    g_{th}\,\dot{t}\,\dot{h} &\propto s^{b(1-z/d)-2+(k-\Delta)} \\
    g_{hh}\,\dot{h}^2 &\propto s^{b(1-z/d)-2+2(k-\Delta)}
\end{align}
In the above equations, the approach of the protocol to the critical point is asymptotically parametrized by $(t,h)\sim(s,s^k)$. This parametrization, subject to $k\ge\Delta$, assures $h/t^\Delta\to const.$ in the limit $s\to 0$. Regardless of the exact value of $k$, the convergence condition becomes
\begin{equation}
    b\!\left(1 - \frac{z}{d}\right) > 0
\end{equation}
Similarly, for $\frac{h}{t^\Delta} \to \infty$,
\begin{equation}
    g_{ij}\dot{\lambda}^i\dot{\lambda}^j \propto s^{p(1-z/d)-2}
\end{equation}
so convergence requires
\begin{equation}
    p\!\left(1 - \frac{z}{d}\right) > 0
\end{equation}
These relations can be used to explore the distance divergence across phase transitions in various models. As summarized in Table~\ref{tab:dynamical-divergences}, the thermodynamic lengths diverge for the 2D Ising and 2D Potts models. Surprisingly, however, it remains finite for the 3D Ising model.

\begin{comment}
    \begin{table}[h]
    \centering
    \begin{tabular}{c|c||c|c|c}
        \multicolumn{2}{c||}{} & $d$ & $z$ & $z/d$ \\
        \hline \hline
        \multirow{2}{*}{2D Ising} & Class A & $2$ & $2.165(10)$ & $1.083(5)$ \\
                                  & Class B & $2$ & $2.235(10)$ & $1.117(5)$ \\
        \hline
        \multicolumn{2}{c||}{3D Ising} & $3$ & $2.032(4)$ & $0.677(1)$ \\
        \hline
        \multicolumn{2}{c||}{2D 3-state Potts} & $2$ & $2.198(2)$ & $1.099(1)$ \\
        \hline
        \multicolumn{2}{c||}{2D 4-state Potts} & $2$ & $2.290(3)$ & $1.095(2)$ \\
    \end{tabular}
    \caption{Dynamical scaling parameters for representative universality classes \cite{odor_universality_2004}.}
    \label{tab:dynamical-divergences}
\end{table}
\end{comment}
\begin{table}[h]
    \centering
    \renewcommand{\arraystretch}{1.5} % Increase vertical padding
    \begin{tabular}{c|c||c|c|c}
        \multicolumn{2}{c||}{} & $b$ & $p$ &  $z/d$ \\
        \hline \hline
        \multirow{2}{*}{2D Ising} & Class A & \multirow{2}{*}{2} & \multirow{2}{*}{$\frac{16}{15}$} & $1.083(5)$ \\
                                  & Class B & & & $1.117(5)$ \\
        \hline
        \multicolumn{2}{c||}{3D Ising} & $1.8899(3)$ & $ 1.2088(1) $ & $0.677(1)$ \\
        \hline
        \multicolumn{2}{c||}{2D 3-state Potts} & $\frac{5}{3}$ & $\frac{15}{14}$ & $1.099(1)$ \\
        \hline
        \multicolumn{2}{c||}{2D 4-state Potts} & $\frac{4}{3}$ & $\frac{16}{15}$ & $1.095(2)$ \\
    \end{tabular}
    \caption{Scaling parameters for representative universality classes \cite{odor_universality_2004}.}
    \label{tab:dynamical-divergences}
\end{table}

\paragraph{Linear Response Applicability:}
As shown in previous works, the validity of the dissipation relation given in Eq.\eqref{Eq:DissipationRelation} requires a separation of timescales between the system’s relaxation rate and the rate of change of the control parameters \cite{rotskoffGeometricApproachOptimal2017}.
This condition can be satisfied in certain models.
Assume that the relaxation time diverges near the phase transition as $\tau\propto (\lambda^\mu)^{-\kappa}$ for some reduced control parameter $\lambda^\mu$ that goes to 0 at the phase transition, and $\kappa>0$. If the protocol is driven with $\dot{\lambda}^\mu \propto (\lambda^\mu) ^ a$ with $a>\kappa$, so that the protocol slows down faster than the response time diverges, then the trajectory is in the linear-response regime  throughout the evolution. Such protocols correspond to algebraic slowdowns near the critical point and are physically realizable \cite{holtzman2025shortcuts}.
To ensure that this is a finite time protocol, we require that the total driving time
\begin{equation}
    T = \intop_0 \frac{d\lambda^\mu}{\dot{\lambda}^\mu} < \infty
\end{equation}
remains finite, which holds for $a<1$. Therefore, finite-time protocols that remain entirely within the linear-response regime exist if and only if $\kappa < 1$. Note that the condition $\kappa<1$ is independent of the geometric convergence conditions derived earlier.

From Eq.~\eqref{Eq:tauPropto}, the two relevant cases are:
\begin{align}
    \frac{h}{t^\Delta} & \to \text{const.} \, : & \lambda^\mu & =t & \kappa & = \frac{bz}{d} \\
     \frac{h}{t^\Delta} & \to \infty \, : & \lambda^\mu & = h & \kappa &= \frac{pz}{d}
\end{align}
Although the examples discussed above do not necessarily satisfy $\kappa < 1$, models with this property do exist.
One notable case is the quantum 2D transverse-field Ising model, which belongs to the same universality class as the 3D classical Ising model \cite{ZHANG2021114632,10.1143/PTP.56.1454}, and hence has finite length across the phase transition \footnote{See \cite{scandi2019thermodynamic} for discussion on thermodynamic geometry in quantum systems}.
Using the corresponding critical exponents from the 3D Ising model and the dynamic critical exponent $z = 1$ \cite{Continentino_2017, Sachdev_2011}, we find \cite{MatthiasVojta_2003}:
\begin{align}
    \frac{h}{t^\Delta} & \to \text{const.} \, : & \kappa & = \frac{bz}{d+z} = 0.630 < 1 \\
     \frac{h}{t^\Delta} & \to \infty \, : & \kappa &= \frac{pz}{d+z} = 0.402 < 1
\end{align}
Thus, the quantum 2D transverse-field Ising model provides an example of a system where the phase transition can be crossed at a finite time, with a finite dissipated availability.

\paragraph{An example:}

The antiferromagnet Ising model in the presence of a uniform magnetic field is a useful example for our analysis, as in these systems, at high enough dimension, it is impossible to traverse from the antiferromagnetic to the disordered phase without crossing a second-order phase transition line \cite{wang1997critical}. Exact solutions for the free energy in antiferromagnetic Ising models in the presence of external field are known only for $1d$ and mean field settings, and in order to explore the crossing of a phase transition we use the latter. With the framework discussed above, we next determine the least-dissipation path between two states 
$\left(T_1,h_1\right)\!\to\!\left(T_2,h_2\right)$, for the mean-field antiferromagnetic model introduced in \cite{vivesUnifiedMeanfieldStudy1997}. 

In this model, the spins $s_i$ are equally split into sublattices $A$ and $B$. A spin in sublattice $A$ interacts with all spins in sublattice $B$ with equal interaction strength, which is antiferromagnetic in sign. There are no interactions between spins in the same sublattice.
Let $m_A=N^{-1}\sum_{i\in A} s_i$ and $m_B = N^{-1}\sum_{i\in B}s_i$ denote the sublattice magnetizations ($N$ is the number of spins in the sublattice). 
The Massieu potential of the system is given by \cite{vivesUnifiedMeanfieldStudy1997}
\begin{eqnarray}
    \psi \;=\ln Z&= -\frac{1}{2}\!\Big[&  \beta K\, m_A m_B \;-\; \beta h\, (m_A+m_B)\nonumber \\
      & &\left. \;+\; \,\big(\vartheta(m_A)+\vartheta(m_B)\big) \right]
\end{eqnarray}
where $\beta=1/T$, $K=-J$ with $J<0$ (antiferromagnetic coupling), and we define
\begin{equation}
    \vartheta(x) \;=\; \frac{1}{2}\Big[(1+x)\log(1+x)+(1-x)\log(1-x)\Big].
\end{equation}

Since there is no notion of correlation length in this model, we use the response time directly to evaluate the relaxation rates. As shown in the end matter, assuming Glauber rates for a single transition with a microscopic timescale $\Gamma$, the dissipation metric is given by 
\begin{widetext}
\begin{equation}
\resizebox{\columnwidth}{!}{$
\begin{aligned}
    g_{\mu\nu}
    \;= & \; \frac{\beta}{2\Gamma\big(1-\beta^2 K^2 \zeta_A \zeta_B\big)^{\!2}}
    \begin{pmatrix}
    \scriptstyle K^2 \!\left[\!\big(1+\beta^2 K^2 \zeta_A \zeta_B\big)\big(m_A^2 \zeta_B + m_B^2 \zeta_A\big) - 4\beta K\, m_A m_B \zeta_A \zeta_B \!\right]
    &
    \scriptstyle -K \!\left[\!\big(1+\beta^2 K^2 \zeta_A \zeta_B\big)\big(m_A \zeta_B+m_B \zeta_A\big) - 2\beta K (m_A+m_B)\zeta_A\zeta_B \!\right]
    \\[6pt]
    \scriptstyle -K \!\left[\!\big(1+\beta^2 K^2 \zeta_A \zeta_B\big)\big(m_A \zeta_B+m_B \zeta_A\big) - 2\beta K (m_A+m_B)\zeta_A\zeta_B \!\right]
    &
    \scriptstyle \big(1+\beta^2 K^2 \zeta_A \zeta_B\big)(\zeta_A+\zeta_B) - 4\beta K\,\zeta_A\zeta_B
    \end{pmatrix}\!
\end{aligned}
$}
\end{equation}
\end{widetext}
where $\zeta_i = 1-m_i^2$, and where $m_A$ and $m_B$ are the minimizer of $\psi$, satisfying the mean-field equations
\begin{align}
    m_A &= \tanh\!\left(\beta h - \beta K\, m_B\right) \label{eq:SCEmAB-A}\\
    m_B &= \tanh\!\left(\beta h - \beta K\, m_A\right) \label{eq:SCEmAB-B}
\end{align}
Let us discuss the implications of this metric on the two sides of the phase transition: the disordered and ordered phases.

\begin{figure}[h]
    \centering
    \includegraphics[width=0.95\linewidth]{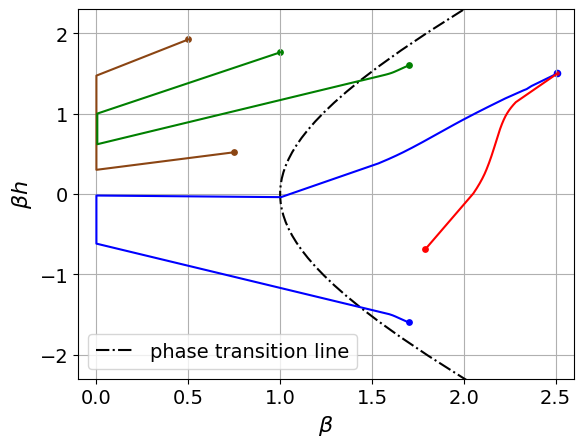}
    \caption{Illustration of the three types of minimal paths in the mean-field antiferromagnet. The dashed black line is the phase transition between the disordered phase in the left and the antiferromagnetic in the right. The colored lines are example for optimal trajectories.}
    \label{fig:shortestPaths}
\end{figure}

In the disordered phase, $m_A=m_B\equiv m$ (hence $\zeta_A=\zeta_B\equiv \zeta$). 
The metric in this case reduces to
\begin{equation}
    g_{\mu\nu} \;=\; \frac{\beta\,\zeta}{\Gamma\bigl(1+\beta K \zeta\bigr)^2}
    \begin{pmatrix}
        K^2 m^2 & -K m \\[2pt]
        -K m & 1
    \end{pmatrix}.
\end{equation}
The metric is rank-one for any value of $m$, with $\det g=0$. This is physically expected: in the disordered phase the system is uniquely defined by a single order parameter -- the total magnetization $m$, therefore there is no dissipation associated with changing the control parameters $\beta$ and $h$ in a combination that does not change $m$. Indeed, the null eigenvector aligns with directions of constant $m$. Geometrically, the least-dissipation trajectory is then the projection of the full geodesic flow onto this one-dimensional submanifold, collapsing the dynamics of the manifold onto its non-degenerate component. The problem of minimizing distances therefore reduces to a 1D minimization: at what distance from the phase boundary does the length minimize?  In this model, the minimum occurs at $\beta=0$ (see End Matter), namely at infinite temperature. This means that the optimal trajectory between two points in this phase follows the null trajectory from the initial point to $\beta=0$, traverse along $\beta=0$, then return along a null trajectory to the final point (see the brown trajectory in Fig.~\ref{fig:shortestPaths}).

In the ordered phase, near the critical line, $m_A$ and $m_B$ can be expanded as shown in \cite{vivesUnifiedMeanfieldStudy1997}:
\begin{align}
    m_{A,B} &\approx \pm\sqrt{1-\frac{1}{\beta_c K}} \;\pm\; \sqrt{\Delta} \\
    \zeta_{A,B} &\approx \frac{1}{\beta_c K}\bigl(1 \pm 2\sqrt{\Delta}\bigr)
\end{align}
where $\Delta$ is a linear combination of $(\beta-\beta_c)$ and $(\beta h -\beta_ch_c)$, namely the linear deviation of the thermodynamic parameters $\beta$ and $\beta h$ from their critical values (see End Matter). 

Substituting these expressions into the metric yields, to leading order,
\[
g_{\mu\nu}\;\propto\;\Delta^{-1}
\]
Hence, while the metric diverges as $\Delta\!\to\!0$, the associated thermodynamic length remains finite.

We identify four distinct classes of minimal (shortest) paths. If both initial and final points are in the disordered phase, then the shortest path is achieved by traversing along the constant magnetization towards $\beta=0$, then changing the external magnetic field until the correct value of total magnetization is achieved, and traversing back along a constant total magnetization line to the final point.

In the second case, the initial and final states are in different phases. The optimal trajectory, which for simplicity we choose to initiate in the antiferromagnetic phase, starts from the initial point to the nearest point on the phase transition line, crosses into the disordered phase, and then ascends to infinite temperature ($\beta = 0$). Once at $\beta = 0$, the trajectory traverses at this value of $\beta$---where the thermodynamic length is minimal---until reaching the final magnetization value, and finally descends along the constant-magnetization line to the final point.

In the third case, both points are in the antiferromagnetic phase, and the geodesic line connecting them is shorter than any path that detours through the disordered region.  Here, the minimal path lies entirely within the ordered manifold.

The fourth case occurs when both endpoints are in the antiferromagnetic phase, but it is nevertheless shorter to cross the phase transition back and forth to the disorder phase than to remain entirely within the ordered region across all the trajectory. Because paths within the disordered phase have zero thermodynamic length except for the section at $\beta=0$, the total length of such a trajectory is often dominated by the two crossings, and can be shorter than any trajectory that does not cross the phase transition.
An illustration of these paths is shown in Fig.~\ref{fig:shortestPaths}.

The thermodynamic lengths within the antiferromagnetic phase were computed using the Fast Marching Method \cite{kimmelComputingGeodesicPaths1998}, following the approach used in \cite{rotskoffDynamicRiemannianGeometry2015}. 
Minimal paths were obtained by performing gradient descent along the numerically computed distance function.

\paragraph{Conclusions} To conclude, we showed that although the dissipation metric diverges at a second order phase transition, for some universality classes crossing the phase transition can be achieved at a finite thermodynamic length. Surprisingly, we find that the optimal trajectory between two points that are in the same phase might even cross to the other phase and back to the initial phase. Many related problems, however, remain open: How can first order phase transitions be incorporate to such a framework? Are there models where the critical exponents vary continuously along the phase transition line (e.g. the 2D Ashkin-Teller model \cite{aoun2024phase}) such that crossing the phase transition cost infinite dissipation for some parts of the phase transition line, but finite for others? Lastly, many heat engines and refrigerators work intentionally near the phase transition \cite{campisi2016power}. Using the suggested framework to optimize their protocol can be an interesting application.   

\paragraph{Acknowledgments} O.R. acknowledges financial support from ISF Grant No. 232/23 and from the Minerva foundation with funding from the Federal German Ministry for Education and Research. We thank David Sivak for many useful discussions and for his hospitality, and Grant Rotskoff for pointing out that the protocol might be too slow to cross the phase transition.

\bibliography{ThermodynamicGeometry}
\newpage

\section{Calculating the dissipation metric}
\begin{comment}
It is convenient to define the total and staggered magnetizations
\begin{align}
    m &= \frac{m_A+m_B}{2}\\
    s &= \frac{m_A-m_B}{2}
\end{align}

For each $m\in[-1,1]$, the corresponding critical parameters $(\beta_c,\alpha_c)$ on the phase boundary satisfy
\begin{align}
    m &= \pm\sqrt{1-\frac{1}{\beta_c K}} \label{m(beta)}\\
    \alpha_c &= \pm\Big(\beta_c K\, m + \arctanh m\Big) \label{alpha(beta,m)}
\end{align}
\end{comment}

%\subsubsection{Calculation of the Metric}
The computation is done as follows: First, we calculate the equilibrium covariances, namely we evaluate 
$\langle \delta m_i(0)\,\delta m_j(0)\rangle$ ($i,j\in\{A,B\}$).
From these, we calculate the time integrated correlations by incorporate linearized relaxation. As these calculations are easier in $m_i$ variables, we then convert to the $\boldsymbol{\lambda}=(\beta,\beta h) \equiv (\beta, \alpha)$ representation, in which the metric is provided. We set $\alpha\equiv \beta h$ for brevity.

Let $\zeta_i\equiv 1-m_i^2$ for $i\in\{A,B\}$. 
The inverse covariance is the Hessian of the function $\psi$ with respect to the coordinates $(m_A,m_B)$. Using the expression for $\psi$, it is given by
\begin{equation}
    -\Sigma^{-1} \;=\; \frac{1}{2}\!
    \begin{pmatrix}
        \dfrac{1}{\zeta_A} & \beta K \\
        \beta K & \dfrac{1}{\zeta_B}
    \end{pmatrix}.
\end{equation}
Its inverse is given by
\begin{equation}
    \Sigma \;=\; \frac{2}{\,1-\beta^2 K^2 \zeta_A \zeta_B\,}
    \begin{pmatrix}
        \zeta_A & -\beta K\,\zeta_A\zeta_B \\
        -\beta K\,\zeta_A\zeta_B & \zeta_B
    \end{pmatrix}
\end{equation}

The relaxation dynamics we use \cite{klichMpembaIndexAnomalous2019}, corresponds to single spin Glauber dynamics, and are given by
\begin{align}
    \dot m_A &= \Gamma\!\left[\tanh\!\left(\alpha - \beta K\, m_B\right) - m_A\right]\\
    \dot m_B &= \Gamma\!\left[\tanh\!\left(\alpha - \beta K\, m_A\right) - m_B\right]
\end{align}
with microscopic attempt rate $\Gamma$. 

Linearizing about equilibrium gives
\begin{align}
    \delta \dot m_A &= \Gamma\left(-\,\delta m_A \;-\; \beta K\,\zeta_A\,\delta m_B\right)\\
    \delta \dot m_B &= \Gamma\left(-\,\beta K\,\zeta_B\,\delta m_A \;-\; \delta m_B\right)
\end{align}
In vector form, 
\begin{eqnarray}
    \begin{pmatrix}
        \delta \dot m_A\\
        \delta \dot m_B
    \end{pmatrix} &=& -\Gamma 
    \begin{pmatrix}
        1 & \beta K\,\zeta_A \\
        \beta K\,\zeta_B & 1
    \end{pmatrix} \begin{pmatrix}
        \delta  m_A\\
        \delta  m_B
    \end{pmatrix} \nonumber\\
    &=& -M \begin{pmatrix}
        \delta m_A\\
        \delta  m_B
    \end{pmatrix}
\end{eqnarray}

%We require
%\begin{equation}
%    \beta \int_{0}^{\infty}\!\langle \delta m_i(0)\,\delta m_j(t)\rangle\,dt
%    \;=\; \beta \int_0^\infty \!\langle y(0)\,y^{\!\top}(t)\rangle\,dt
%\end{equation}
Since $\langle \delta m_i(0)\,\eta(t)\rangle=0$ for $t>0$, substituting $\delta \vec m(t)=e^{-Mt}\delta \vec m(0)$ yields
\begin{widetext}
\begin{equation}
\begin{aligned}
    \beta \int_0^\infty \!\langle \delta m_i(0)\,\delta m_j (t)\rangle\,dt
    &= \beta\,\langle \delta m_i(0)\,\delta m_j (0)\rangle \int_0^\infty \!e^{-M^{\!\top}t}\,dt
     \;=\; \beta\,\Sigma\,(M^{\!\top})^{-1} \\[3pt]
    &= \frac{2\beta}{\Gamma\big(1-\beta^2 K^2 \zeta_A \zeta_B\big)^{\!2}}
    \begin{pmatrix}
        \zeta_A\big(1+\beta^2 K^2 \zeta_A \zeta_B\big) & -\,2\beta K\,\zeta_A\zeta_B\\
        -\,2\beta K\,\zeta_A\zeta_B & \zeta_B\big(1+\beta^2 K^2 \zeta_A \zeta_B\big)
    \end{pmatrix}\!
\end{aligned}
\end{equation}
\end{widetext}

To express the dissipation metric in the control variables $\boldsymbol{\lambda}=(\beta,\beta h)$, we transform fluctuations to $(-E,m)$ via
\begin{align}
    E &= \tfrac{1}{2} K\, m_A m_B\\
    m &= \tfrac{1}{2}(m_A+m_B)
\end{align}
so the Jacobian is
\begin{equation}
    \mathcal{J} \;=\; \frac{\partial(-E,m)}{\partial(m_A,m_B)} 
    \;=\; \frac{1}{2}
    \begin{pmatrix}
        -K m_B & -K m_A \\
        \;1 & \;1
    \end{pmatrix}
\end{equation}
The dissipation metric then reads
\begin{widetext}
\begin{equation}
\resizebox{\columnwidth}{!}{$
\begin{aligned}
    g_{\mu\nu}
    \;= & \; \beta\, \mathcal{J}\,\Sigma\,(M^{\!\top})^{-1} \mathcal{J}^{\!\top} \\
    \;= & \; \frac{\beta}{2\Gamma\big(1-\beta^2 K^2 \zeta_A \zeta_B\big)^{\!2}}
    \begin{pmatrix}
    \scriptstyle K^2 \!\left[\!\big(1+\beta^2 K^2 \zeta_A \zeta_B\big)\big(m_A^2 \zeta_B + m_B^2 \zeta_A\big) - 4\beta K\, m_A m_B \zeta_A \zeta_B \!\right]
    &
    \scriptstyle -K \!\left[\!\big(1+\beta^2 K^2 \zeta_A \zeta_B\big)\big(m_A \zeta_B+m_B \zeta_A\big) - 2\beta K (m_A+m_B)\zeta_A\zeta_B \!\right]
    \\[6pt]
    \scriptstyle -K \!\left[\!\big(1+\beta^2 K^2 \zeta_A \zeta_B\big)\big(m_A \zeta_B+m_B \zeta_A\big) - 2\beta K (m_A+m_B)\zeta_A\zeta_B \!\right]
    &
    \scriptstyle \big(1+\beta^2 K^2 \zeta_A \zeta_B\big)(\zeta_A+\zeta_B) - 4\beta K\,\zeta_A\zeta_B
    \end{pmatrix}\!
\end{aligned}
$}
\end{equation}
\end{widetext}

%Figures~\ref{fig:sivakmetric_antiferro} show the resulting metric components.
\section{Minimal Dissipation in Disordered Phase}
In the disordered phase, the magnetizations of the two sublattices are identical. Therefore, the total magnetization
\begin{align}
    m &= \frac{m_A+m_B}{2} = m_A = m_B
\end{align}

For each $m\in[-1,1]$, the corresponding critical parameters $(\beta_c,\alpha_c)$ on the phase boundary satisfy 
\begin{align}
    m &= \pm\sqrt{1-\frac{1}{\beta_c K}} \label{m(beta)}\\
    \alpha_c &= \pm\Big(\beta_c K\, m + \arctanh m\Big) \label{alpha(beta,m)}
\end{align}

\paragraph{Eigenstructure and the zero-length direction.}
The eigenpairs of the bracketed matrix are
\begin{align}
    \lambda_1 &= 0, 
    & \boldsymbol{v}_1 &= \begin{pmatrix} 1 \\ K m \end{pmatrix},
    \\
    \lambda_2 &= 1+K^2 m^2, 
    & \boldsymbol{v}_2 &= \begin{pmatrix} K m \\ -1 \end{pmatrix}.
\end{align}
Thus the zero mode $\boldsymbol{v}_1$ follows lines of constant $m$ in $(\beta,\alpha)$.

\paragraph{Scaling of the metric prefactor.}
From \eqref{m(beta)} we have $\zeta=1-m^2=1/(\beta_c K)$ along the phase boundary labeled by magnetization $m$, with critical parameter $\beta_c$.
Using this in the prefactor gives
\begin{equation}
    g_{\mu\nu} \;\propto\; 
    \frac{\beta}{\bigl(\beta_c+\beta\bigr)^2}\,\bigl(1+K^2 m^2\bigr).
\end{equation}

\paragraph{Scaling of the displacement along the zero mode.}
Consider two neighboring constant-$m$ lines labeled by 
$(\beta_{c1},\alpha_{c1})$ and $(\beta_{c2},\alpha_{c2})=(\beta_{c1}+\delta_\beta,\alpha_{c1}+\delta_\alpha)$.
We find their intersection with a straight line parallel to the zero mode at each point:
\begin{align}
    \begin{pmatrix}\beta\\ \alpha\end{pmatrix}
    &= a\,\boldsymbol{v}_1(\beta_{c1}) + \begin{pmatrix}\beta_{c1}\\ \alpha_{c1}\end{pmatrix}
    \label{eq:b0a0-1}\\[2pt]
    \begin{pmatrix}\beta\\ \alpha\end{pmatrix}
    &= b\,\boldsymbol{v}_1(\beta_{c2}) + \begin{pmatrix}\beta_{c1}+\delta_\beta\\ \alpha_{c1} + \dfrac{\partial \alpha_c}{\partial \beta_c}\,\delta_\beta \end{pmatrix}
    \label{eq:b0a0-2}
\end{align}
Subtracting \eqref{eq:b0a0-1} from \eqref{eq:b0a0-2} yields
\begin{equation}
    \delta_\beta 
    \begin{pmatrix} 1 \\[2pt] \dfrac{\partial \alpha_c}{\partial \beta_c} \end{pmatrix}
    =
    \begin{pmatrix}
        1 & 1 \\
        K m_2 & K m_1
    \end{pmatrix}
    \begin{pmatrix} -b \\[2pt] a \end{pmatrix}
\end{equation}
where $m_1=m(\beta_{c1})$, $m_2=m(\beta_{c2})$.
From \eqref{alpha(beta,m)} we have 
\(
\dfrac{\partial \alpha_c}{\partial \beta_c} = \dfrac{1}{m}
\)
Inverting gives
\begin{equation}
\begin{aligned}
    a \;&=\; \frac{1-m^2}{K m}\,\frac{\partial m}{\partial \beta_c}
      \;=\; \frac{1-m^2}{K m}\,\frac{2m}{(1-m^2)^2}\\
      \;&=\; \frac{2}{K(1-m^2)}
      \;=\; 2\beta_c
\end{aligned}
\end{equation}
where we used $1-m^2=1/(\beta_c K)$ and $\dfrac{\partial m}{\partial \beta_c} = \dfrac{2m}{(1-m^2)^2}$ from \eqref{m(beta)}.
Hence the displacement along the zero mode scales as 
\(
\delta\boldsymbol{\lambda}\sim (\beta+\beta_c).
\)

\paragraph{Minimum at $\beta=0$.}
Combining the prefactor and displacement scalings,
\[
\sqrt{\delta\lambda^\mu\, g_{\mu\nu}\, \delta\lambda^\nu}
\;\propto\;
\sqrt{\beta}
\]
so the infinitesimal thermodynamic length is minimized at $\beta=0$ (infinite temperature).
Thus, within the disordered phase, the least-dissipation path approaches the high-temperature limit.

\end{document}